\documentclass[prper,aps,twocolumn,groupedaddress,floats,showpacs,final]{revtex4-1}
\usepackage{leading}
\usepackage[latin3]{inputenc}
\usepackage[makeroom]{cancel}
\usepackage{graphicx}
\usepackage{amsmath}
\usepackage{amsfonts}
\usepackage{amssymb}

\usepackage{color}
\usepackage[left=2cm,right=2cm,top=2cm,bottom=2cm]{geometry}
\usepackage{graphicx} 
\usepackage{dcolumn}
\usepackage{bm}
\usepackage{simplewick}
\usepackage{array}
\usepackage{appendix}
\usepackage{cancel}
\usepackage[normalem]{ulem} 
\RequirePackage[
   hyperindex,colorlinks,bookmarksnumbered,
   plainpages=true,pdfstartview=FitH]{hyperref}
\hypersetup{linkcolor=blue,urlcolor=blue,citecolor=blue}
\usepackage{hyperref}
\usepackage{mathrsfs}

\newcommand{\DK}[1]{{\color{black}{#1}}} % Deepak
\newcommand{\DE}[1]{{\color{black}{#1}}} % Deepak
\newcommand{\AP}[1]{{\color{black}{#1}}} % Andrei
\newcommand{\pdag}{{\phantom{\dagger}}}
 \renewcommand{\emph}[1]{\textit{#1}}

\newcommand{\Tk}{T_{\rm K}}

\definecolor{darkblue}{rgb}{0,0,0.5}
\definecolor{darkgreen}{rgb}{0,0.5,0}
\definecolor{darkred}{rgb}{.7,0,0}
\definecolor{purple}{rgb}{0.5,0,0.6}
\definecolor{orange}{rgb}{1,0.5,0}
\definecolor{grey}{rgb}{.6,.6,.6}
\definecolor{lightpink}{rgb}{1,0.7,0.75}
\definecolor{pink}{rgb}{1,0.4,0.58}
\definecolor{deeppink}{rgb}{1,0.08,0.58}
\usepackage{upgreek}

\begin{document}
\title{Multistage Kondo effect in a multiterminal geometry: A modular quantum interferometer}

\author{D. B. Karki}
\affiliation{Division of Quantum State of Matter, Beijing Academy of Quantum Information Sciences, Beijing 100193, China}
\author{Andrei I. Pavlov and Mikhail N. Kiselev}
\affiliation{The Abdus Salam International Centre for Theoretical Physics (ICTP), Strada Costiera 11, I-34151 Trieste, Italy}

\begin{abstract}
Quantum systems characterized by an interplay between several resonance scattering channels demonstrate
very rich physics. To illustrate it we consider a multistage Kondo effect in nanodevices as a paradigmatic model for a multimode resonance scattering. We show that the channel crosstalk results in a destructive interference between the modes. This interplay can be controlled by manipulating the tunneling junctions in the multilevel and multiterminal geometry. We present a full-fledged theory of the multistage Kondo effect at the strong-coupling Fermi-liquid fixed point and discuss the influence of quantum interference effects to the quantum transport observables.
\end{abstract}

\date{\today}
\maketitle

%%%%%%%%%%%%%%%%%%%%%%%%%%%%%%%%%%%%%%% TITLE %%%%%%%%%%%%%%%%%%%%%%%%%%%%%%%%%%%%
\textit{Introduction.}
\DE{The exchange coupling between a localized spin and conduction electrons at low temperature gives rise to \AP{the} Kondo screening phenomenon\AP{~\cite{Kondo, Nozieres,Hewson}. This phenomenon has been extensively studied over decades and often serves as a test bed of strongly correlated physics~\cite{Cox_Adv_Phys(47)_1998}}. Depending on the size of the localized spin $\mathcal{S}$ and the number of conduction channels $\mathcal{K}$, the ground state of the system falls into one of three classes, often referred as fully screened ($\mathcal{K}{=} 2\mathcal{S}$), underscreened ($\mathcal{K}{<} 2\mathcal{S}$), and overscreened ($\mathcal{K}{>} 2\mathcal{S}$) cases~\cite{Nozieres_Blandin_JPhys_1980}. Among them, the fully screened and underscreened Kondo effects are completely described by a local Fermi-liquid (FL) theory~\cite{Nozieres}.} \AP{Recently, remarkable progress was achieved in controllable realizations of various fully screened Kondo phenomena in nanostructures~\cite{new, chines_2sk}, further fuelling the continued interests in this field. With the two-stage Kondo effect being now a subject of experimental studies, it is now a question if more general multiterminal setups can contain principally new and richer physics  in comparison to the currently realized ones.}

\DE{The fully screened Kondo effect, although described by FL theory, \AP{possesses} several exotic properties \AP{beyond} its trivial generalization of a single channel, corresponding to $\mathcal{S}=1/2$ and $\mathcal{K}=1$~\cite{Cox_Adv_Phys(47)_1998, lastref2}.} The $\mathcal{S}=1/2$ Kondo effect is unlikely to be sufficient for the
complete description of the physics of a magnetic impurity
in a nonmagnetic host since the truncation
of the impurity spectrum to one level is not possible~\cite{Glazman_PRL_2001}. \AP{Thus,} the consistent description requires the consideration of several orbitals of conduction electrons $\mathcal{K}>1$
which interact with the higher-spin $\mathcal{S}>1/2$ of the localized
magnetic impurity~\cite{intspin, Wiel_PRL(88)_2002, depl}. The recent experiment~\cite{chines_2sk} has further shed light on the relevance of high-spin Kondo effects in nanostructures. In addition, with the rapid progress of semiconductor quantum dot technologies, the understanding and
control over the high-spin state properties have been quickly expanding in recent years with the ambition of
using high-spin states for quantum information processing~\cite{tarucha_2021}.

\begin{figure}[t]
\includegraphics[scale=0.22]{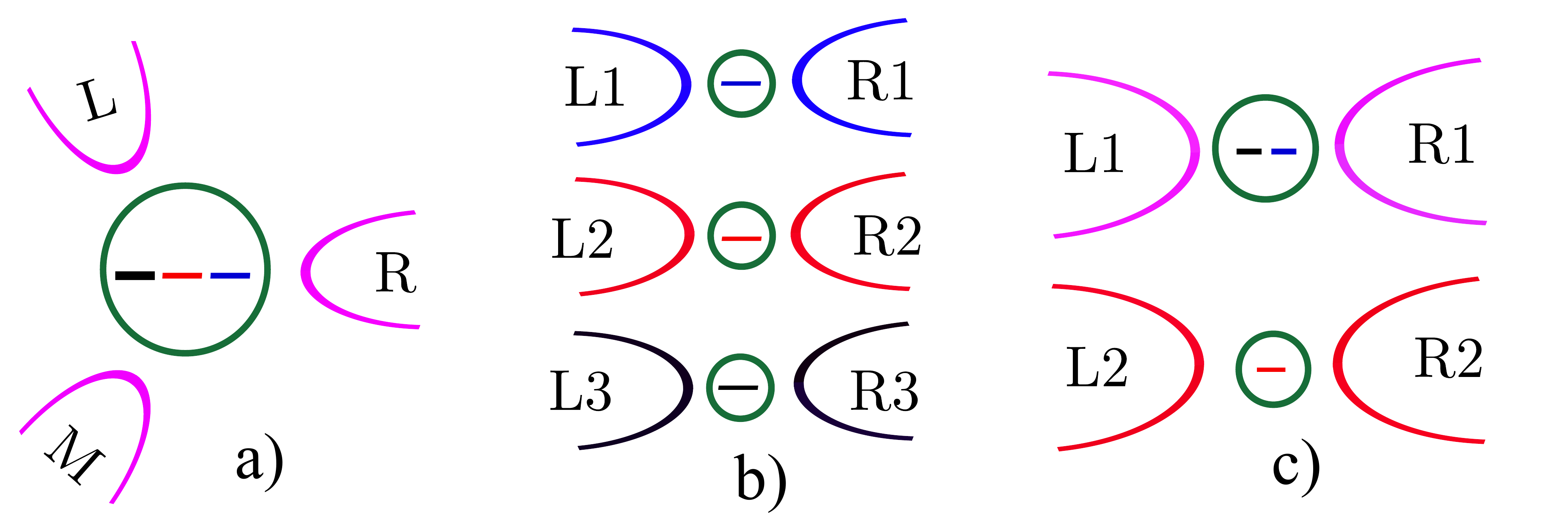}
\includegraphics[width=55mm]{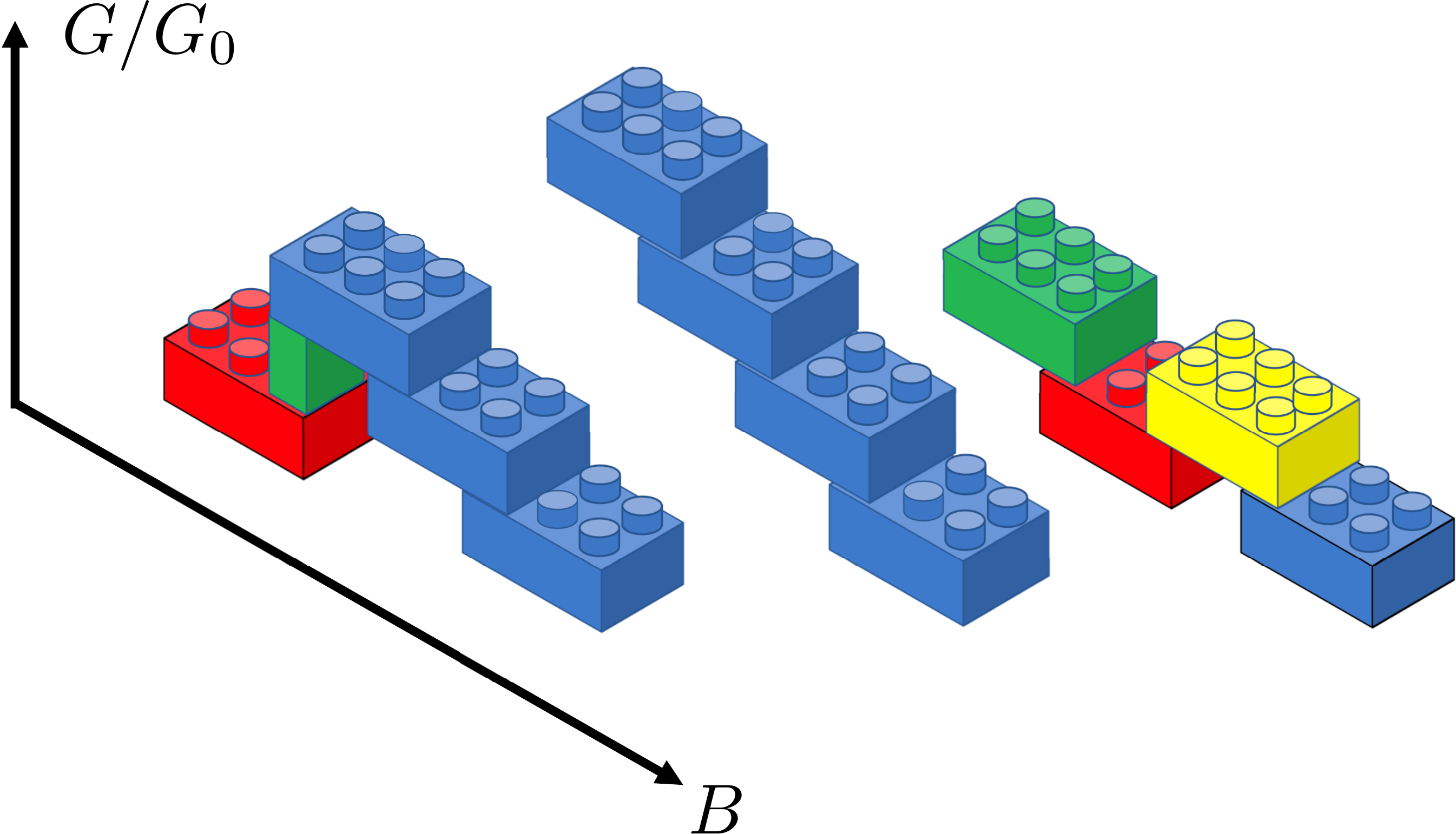}
\vspace*{-3mm}
\caption{{\color{black}Top:} The schematic representation of three-channels $\mathcal{S}=3/2$ Kondo setups studied in this Letter. (a) The series configuration exhibiting 3SK effects via three interfering channels, (b) the capacitively coupled system forming a parallel configuration for $\mathcal{S}=3/2$ Kondo effects with decoupled resonant channels, and (c) a hybrid Kondo setup allowing for the study of the interplay between multistage and single-stage Kondo effects (see text for details). Bottom:
Cartoon for the schematic dependence of the differential conductance
on a magnetic field for (a)$-$(c) setups. Blue ``ascending" modular
blocks denote different stages of Kondo screening. Red ``descending"
modular blocks stand for effects of destructive interference. Yellow
and green modular blocks are used to highlight the nonmonotonicity.}\label{cond_1}
\end{figure}

The prototypical example of multichannel fully screened Kondo effects corresponds to the case of $\mathcal{S}=1$ impurity coupled to $\mathcal{K}=2$ conduction channels. It is well known that the two-terminal (2T) setup offers\AP{, as maximum,} two distinct channels built as the linear combinations (symmetric and antisymmetric) of electron states in the left (L) and right (R) terminal.  These channels will be referred to without loss of generality as even and odd channels, respectively~\cite{GR}. Based on the ways of how to realize $\mathcal{K}=2$, two different cases of the $\mathcal{S}=1$ Kondo effect emerge. The first case corresponds to a 2T realization \AP{with} an explicit coupling between the even and odd channels in order to have $\mathcal{K}=2$. The \AP{other case} is achieved by using capacitively coupled four terminals (two pairs of left and right leads) \AP{to provide} two Kondo channels (one channel from each pair of terminals) necessary for the screening of the $\mathcal{S}=1$ impurity~\cite{aak1, aak2, Mora_Schuricht_(89)_2014, twoc, dee3, mkadd}. While the strong-coupling regime of \AP{the} former case results in completely destructive interference~\cite{twoc}, the latter one allows us to have fully constructive interferences~\cite{aak1}. 
These two cases are commonly referred to as series and parallel configurations of two-stage Kondo effects and are detailed in Fig.~\ref{cond_1} (top). The series configuration results in completely destructive interference due to the competition between two Kondo channels, both being at resonant scattering (even and odd channels characterized by respective Kondo temperatures  $\Tk^{\rm e}$ and $\Tk^{\rm o}$). This configuration is known to possess non-monotonic conductance [(Fig.~\ref{cond_1} (bottom)] - a benchmark property for observations of two-stage Kondo (2SK) effects~\cite{Glazman_PRL_2001}.

Multi ($N{>}2$)-terminal nanodevices have attracted great attention from both theoretical and experimental communities for their potential use in nanotechnologies~\cite{ch,casti, lastref1, mk}. In addition, probing several hallmarks of strongly correlated electron systems, such as the Kondo density of states, requires a setup beyond 2T~\cite{3tk}. Likewise, the experimental detection of the Hanbury Brown - Twiss (HBT) correlations requires a minimal setup of 3T geometry~\cite{hanb1}. Moreover, certain classes of Kondo effects, such as the topological Kondo effects, are intrinsically multiterminal effects~\cite{beri}. \DE{Interestingly, the physics of the $\mathcal{S}=1/2$ Kondo impurity coupled to $N$ terminals can be reached by mapping it to the corresponding two-terminal situation since only the even channel is coupled to the impurity~\cite{hanb}}. In contrast, the $N$-terminal Kondo effect with $\mathcal{S}=N/2$ exhibiting a fullyscreened ground state couples all $N=\mathcal{K}$ conduction modes to the impurity degrees of freedom resulting in multiresonant Kondo screening phenomena~\cite{twoc}.

We now turn our attention to the simplest situation of $\mathcal{S}=3/2$ and $\mathcal{K}=3$. While the corresponding parallel configuration needs six terminals, the series setup requires three terminals (3T). In addition, it is evident that the higher-spin parallel configuration is a trivial generalization of the corresponding $\mathcal{S}=1/2$ situation~\cite{Mora_Schuricht_(89)_2014}. Our focus would thus be on the series configuration where the non-trivial interplay between three Kondo channels provides far richer physics over the widely studied \DE{two-stage Kondo effects}. \DE{In this case, one symmetric mode (e) and two modes orthogonal to the even mode (o1 and o2) compete with each other to screen the localized spin $\mathcal{S}=3/2$}. These three channels are characterized by three Kondo temperatures which can be tuned to satisfy a certain hierarchy $\Tk^{\rm e}\geq\Tk^{\rm o1}\geq \Tk^{\rm o2}$ (see below). The fully screened Kondo ground state, the Kondo singlet, would then result from three different stages of screening $\mathcal{S}=3/2\to 1\to 1/2\to 0$. We refer to this phenomenon as the three-stage Kondo (3SK) effect and concentrate our work on the development of low-energy FL theory of transport through the 3SK effect.

\textit{Formulation of problem.} We consider a multilevel quantum impurity (dot) with an effective spin $\mathcal{S}$ coupled to $N$ external terminals. The high-spin state of the dot is achieved by the Hund's coupling (see Ref.~\cite{kka} for details) in the presence of an external Zeeman field $B$. This system is represented by the generic Anderson model~\cite{and1, and2}
\begin{align}\label{model}
H&=\sum_{k\alpha\sigma}\left(\epsilon_k +\varepsilon_\sigma^Z\right) C^\dagger_{\alpha k \sigma}C_{\alpha k \sigma}+\sum_{\alpha k i \sigma} t_{\alpha i} C^\dagger_{\alpha k \sigma} d_{i \sigma} + \text{H.c.}\nonumber\\
&+\sum_{i\sigma}
(\varepsilon_i +\varepsilon_\sigma^Z) d^\dagger_{i \sigma}d_{i \sigma}
+E_c \hat{\cal N}^2- {\cal I} \hat{\mathcal{S}}^2,
\end{align}
where $C_{\alpha k\sigma}$ annihilates an electron at the terminal $\alpha$ from the momentum state $k$ with spin $\sigma$ ($=\uparrow,\downarrow$) and $\epsilon_k=\varepsilon_k-\mu$
is the energy of conduction electrons with respect to
the chemical potential $\mu$. The electron in the $i$th orbital of the quantum dot with energy $\varepsilon_i+\varepsilon_\sigma^Z$ ($\varepsilon_\sigma^Z=-\sigma B/2$) is described by the operator $d_{i\sigma}$ such that $\hat{\cal N}=\sum_{i\sigma} d^{\dagger}_{i\sigma}d_{i\sigma}$ 
represents the total number of electrons in the dot. The exchange integral accounting for the Hund's rule is represented by $\mathcal{I}$, $E_c$ is the charging energy such that ${\cal I}\ll E_c$, and $t_{\alpha i}$ are the tunneling matrix elements (for details see Refs.~\cite{Coleman_PRL(94)_2005, Coleman_PRB(75)_2007, Coleman_book}).

\textit{Achieving multiple resonant Kondo channels.} As the relevant case of the 3SK effects, we concentrate our discussion on the particular case of a three-level impurity ($i=1, 2, 3$) tunnel-coupled to three external leads: left (L), middle (M) and right (R) (see Fig. \ref{cond_1}). Assuming a total of three electrons, the presence of Hund's coupling results in the quartet configuration of the impurity possessing an effective spin $\mathcal{S}=3/2$. We note that the spin-3/2 quartet state is well separated from the corresponding spin-1/2 doublets [see Supplemental Material (SM)~\cite{suppl} for details]. We then apply the Schrieffer-Wolff  transformation~\cite{Schrieffer_Wolf_PR(149)_1966} to Eq.~\eqref{model} which eliminates the charge fluctuations between the orbitals resulting in the effective Hamiltonian as
\DK{\begin{equation}\label{schriffer}
\mathscr{H}=\mathscr{H}_0+\frac{1}{2}\sum_{\alpha,\alpha'=1}^3\;\sum_{ k\sigma, k'\sigma'}\mathcal{J}_{\alpha\alpha'}{\bf \mathcal{S}}\cdot C^{\dagger}_{\alpha k\sigma}\tau_{\sigma\sigma'}C_{\alpha' k'\sigma'},
\end{equation}}
with $\mathscr{H}_0=\sum_{\alpha k\sigma}\epsilon_{k}\;C^{\dagger}_{\alpha k\sigma}C_{\alpha k\sigma}$ and $\tau_{\sigma\sigma'}$ being the Pauli matrix. The $3\times 3$ Hermitian matrix $\mathbb{J}$ of exchange couplings $\mathcal{J}_{\alpha\alpha'}$ can be expressed in terms of the size of the effective spin $\mathcal{S}$, charging energy $E_c$, and nine complex tunneling elements $t_{\alpha i}$ such that
\begin{equation}
\mathcal{J}_{\alpha\alpha'}=\frac{2}{\mathcal{S}E_c}\sum_{i=1}^3 t^*_{\alpha i}t_{\alpha' i}.
\end{equation}
The $3{\times} 3$ matrix $\mathbb{J}$ possesses at most three non-zero eigenvalues each representing distinct conduction channels~\cite{GP_Review_2005}, let us say $\mathcal{J}_{1, 2, 3}$. To achieve $\mathcal{J}_{1, 2, 3}{>} 0$, the matrix $\mathbb{J}$ must possess the following three invariants $\mathscr{M}_{1, 2, 3}$ such that
\begin{align}
&\mathscr{M}_1={\rm Tr}\;\mathbb{J} =\sum_{i=1}^3\mathcal{J}_i> 0,\;\;\mathscr{M}_2={\rm Det}\;\mathbb{J}=\prod_{i=1}^3\mathcal{J}_i> 0,\nonumber\\
&\mathscr{M}_3=\frac{1}{2}\Big[\left({\rm Tr}\;\mathbb{J}\right)^2{-}{\rm Tr}\;\mathbb{J}^2\Big]=\mathcal{J}_1\mathcal{J}_2{+}\mathcal{J}_2\mathcal{J}_3{+}\mathcal{J}_1\mathcal{J}_3> 0,\nonumber
\end{align}
where ``${\rm Tr}$" and ``${\rm Det}$" stand for the trace and determinant, respectively. The simplest case arises when all $t_{\alpha i}$ are tuned to be equal where $\mathbb{J}$ permits only one eigenvalue, with the other two being zero since for this case $\mathscr{M}_2=0$. The resulting situation describes the single-channel underscreened ($\mathcal{S}>\mathcal{K}=1$) Kondo effects characterized by the channel corresponding to the single nonzero eigenvalue of $\mathbb{J}$.

\begin{figure}
\includegraphics[scale=0.3]{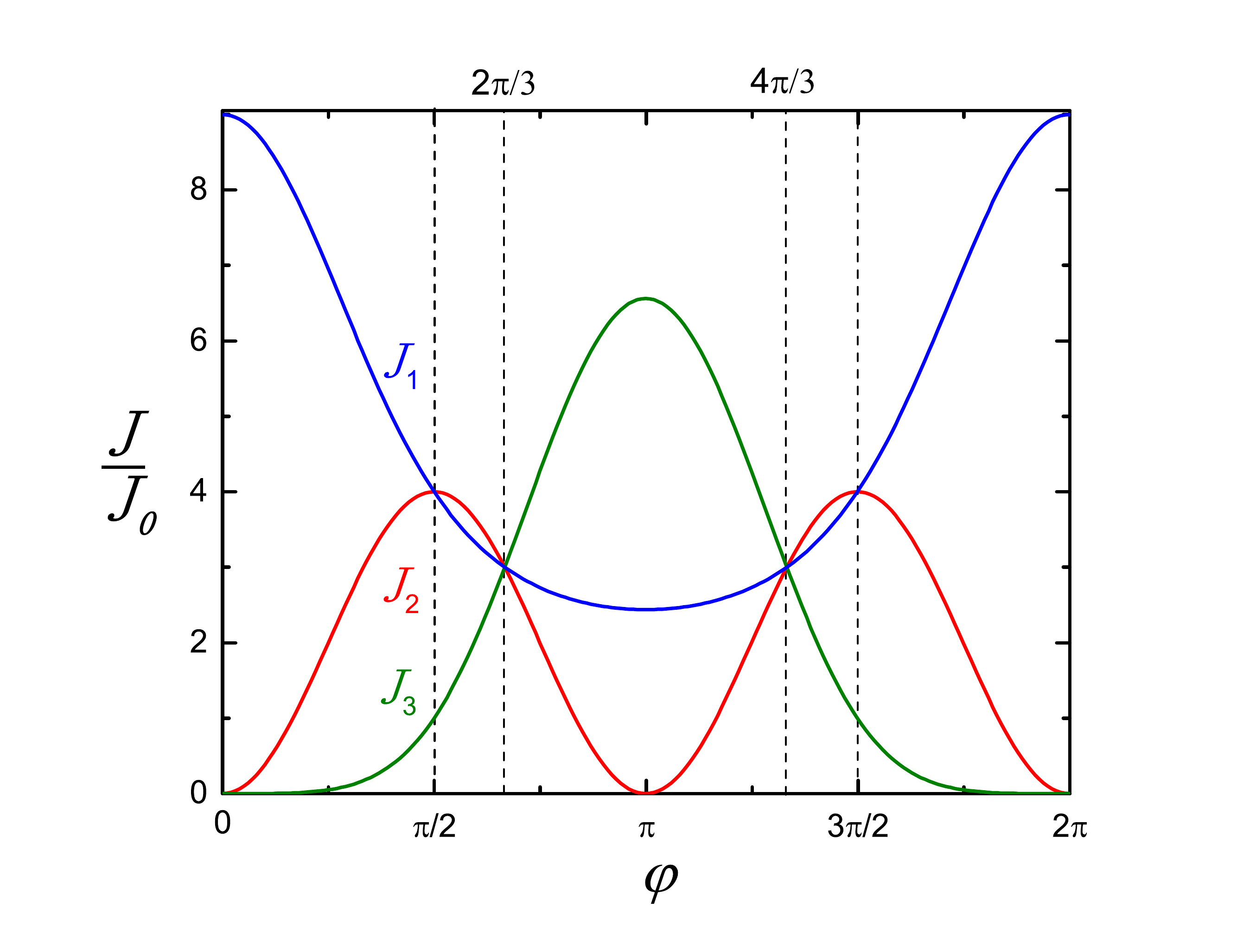}
\vspace*{-7mm}
\caption{The phase tunable eigenvalues of exchange matrix $\mathbb{J}$ providing three distinct Kondo channels (see text for details).}\label{kvv}
\end{figure}

The complex tunneling parameters $t_{\alpha i}=\left|t_{\alpha i}\right|e^{i\varphi_{\alpha i}}$ provide a very large parameter space and hence various $\mathbb{J}$ matrices can be formed. For a simple demonstration of the 3SK effect, one needs all three eigenvalues of $\mathbb{J}$ to be nonzero and positive definite since these eigenvalues will provide independent conduction channels for the Kondo screening. One of the simplest ways to achieve such a goal is to consider a single phase $\varphi\neq 0$ keeping all tunneling amplitudes the same, satisfying the condition $\mathscr{M}_{1, 2, 3}> 0$. To this end, we chose a prototypical realization such that the tunneling elements are parametrized as~\cite{suppl}
\begin{align}\label{tunnelingm}
&t_{\rm L1}=|t|e^{\textit{i}\varphi},\;\; t_{\rm L2}=|t|,\;\; t_{\rm L3}=|t|e^{-\textit{i}\varphi},\nonumber\\
& t_{\rm M1}=|t|e^{-\textit{i}\varphi},\;\; t_{\rm M2}=|t|,\;\; t_{\rm M3}=|t|e^{\textit{i}\varphi},\nonumber\\
& \;\;\;\;\;\;\;\;\;t_{\rm R1}=t_{\rm R2}=t_{\rm R3}=|t|.
\end{align}
This choice results in three eigenvalues $\mathcal{J}_{1,2 , 3}$ of $\mathbb{J}$ matrix which can be tuned with phase $\varphi$ 
\begin{align} 
\mathcal{J}_{1,3} &=\frac{\mathcal{J}_0}{2}\left(7{+}2\cos2\varphi{\pm} \sqrt{32\cos\varphi+(5+2\cos2\varphi)^2} \right),\nonumber\\
&\;\;\;\;\;\;\;\;\;\;\;\;\;\;\;\mathcal{J}_2 =4\mathcal{J}_0\sin^2\varphi,
\end{align}
where we denoted $\mathcal{J}_0=\frac{2|t|^2}{\mathcal{S}E_c}$. It is seen from the above equation that $\varphi=0$ results in $\mathcal{J}_1{>} 0$ with $\mathcal{J}_{2,3}=0$. Thus, by tuning the phase $\varphi\in [0, 2\pi)$, one can explicitly achieve the desirable condition $\mathcal{J}_{1, 2, 3}{>} 0$ which gives three distinct Kondo channels, see Fig.~\ref{kvv}. In addition, tuning $\varphi$ further offers some interesting points where two eigenvalues are equal as well as all of them being equal, keeping the condition $\mathcal{J}_{1, 2, 3}\neq 0$. This provides a viable way of manipulating the strength of the conduction channels. In addition, as detailed in SM~\cite{suppl}, \DK{$\mathcal{J}_{1, 2, 3}{>} 0$} can be achieved even with $\varphi=0$ (the real tunneling elements) by considering asymmetry among the tunneling amplitudes. Therefore, in the following discussion we assume
without loss of generality that all three eigenvalues of the matrix
$\mathbb{J}$ are positive and hence all scattering channels interact with the quantum impurity.

\textit{Rotation of electron states in the leads.}
 The usual situation of 2T geometry always suggests the rotation of electron states given by the Glazman-Raikh (GR) transformation~\cite{GR} [for symmetric coupling it writes $C_{\rm e}\equiv\left(C_{\rm L}+C_{\rm R}\right)/\sqrt{2}$ and $C_{\rm o}\equiv\left(C_{\rm L}-C_{\rm R}\right)/\sqrt{2}$], which paves the way of writing Hamiltonian Eq.~\eqref{schriffer} in diagonal form and hence resulting in a Kondo Hamiltonian. Increasing the number of terminals $N$ but keeping only one nonzero eigenvalue of $\mathbb{J}$ is the trivial case since only the symmetric mode $C_{\rm e}=\sum_{\rm \alpha}C_\alpha/\sqrt{N}$ would be coupled to the impurity, with all other $N-1$ modes orthogonal to $C_{\rm e}$ remaining completely decoupled. Therefore for $N$-terminal geometry with all eigenvalues of $\mathbb{J}$ being nonzero, a transformation similar to GR is more involved. Interestingly, the remaining $N-1$ rotated states which are orthogonal to $C_{\rm e}$ can be formed by using the $N-1$ Cartan generators of the SU$(N)$ group~\cite{lastref1, lastref2, apadd}. For the particular case of 3T geometry, the remaining two rotated states are formed by using the generators of the {\color{black} Cartan} basis of the SU$(3)$ representation, namely $\lambda_3/\sqrt{2}$ and $\lambda_8/\sqrt{2}$.

A naive rotation transformation constructed from $C_{\rm e}$, $C_{\rm o1}\leftrightarrow \lambda_3/\sqrt{2}$, and $C_{\rm o2}\leftrightarrow \lambda_8/\sqrt{2}$~\cite{georgi},
\begin{align}
\begin{pmatrix}
C_{\rm e} \\
C_{\rm o1}\\
C_{\rm o2}
\end{pmatrix}=
\mathbb{U}_\lambda \begin{pmatrix}
C_{\rm L}\\ 
C_{\rm M}\\
C_{\rm R}
\end{pmatrix},\;\mathbb{U}_\lambda=\begin{pmatrix}
\frac{1}{\sqrt{3}} & \frac{1}{\sqrt{3}} & \frac{1}{\sqrt{3}} \\
-\frac{1}{\sqrt{2}} &  \frac{1}{\sqrt{2}} & 0 \\
-\frac{1}{\sqrt{6}}& -\frac{1}{\sqrt{6}} & \frac{2}{\sqrt{6}}
\end{pmatrix},
\end{align}
would neither contain the coupling asymmetry $|t_{\alpha i}|\neq|t|_{\alpha' i'}$ nor the information about the tunneling phase $\varphi\neq 0$. We thus find a very general transformation $\mathbb{U}$ (similar to GR rotation in 2T geometry) for 3T Kondo geometry which provides all eigenvalues of $\mathbb{J}$ to be \DK{positive non-zeros $\mathcal{J}_{1, 2, 3}{>} 0$} by accounting for the tunneling asymmetry and phase. We relegate the discussion of this general transformation to the SM~\cite{suppl} and concentrate our discussion here to the case of having the same tunneling amplitudes $|t|_{\alpha i}$ but with $\varphi\neq 0$. As detailed in the SM, the most general transformation accounting for the choice of Eq.~\eqref{tunnelingm} reads
\begin{align}\label{unitarym}
&\mathbb{U}[\mathcal{C}(\varphi)]=\frac{2}{\sqrt{8+\left(\mathcal{C}+2\right)^2}}\times\overline{\mathbb{U}}[\mathcal{C}(\varphi)],
\end{align}
with
\begin{align}
&\overline{\mathbb{U}}=\left(\!\!\!\!\!
\begin{array}{ccc}
 1 & 1 & 1{+}\frac{\mathcal{C}}{2} \\
{-}\sqrt{\frac{3}{2}}\sqrt{1{+}\frac{\mathcal{C}}{12}\left(\mathcal{C}{+}4\right)} & \sqrt{\frac{3}{2}}\sqrt{1{+}\frac{\mathcal{C}}{12}\left(\mathcal{C}{+}4\right)} & 0 \\
 -\frac{1}{\sqrt{2}}\left(1{+}\frac{\mathcal{C}}{2}\right) & -\frac{1}{\sqrt{2}}\left(1{+}\frac{\mathcal{C}}{2}\right) & \sqrt{2} \\
\end{array}
\!\right),\nonumber\\
&\mathcal{C}(\varphi)=\!\!\sqrt{11{-}4 \cos (\varphi ){+}2 \cos (2 \varphi )}{-}1{-}2 \cos (\varphi ),\;\mathcal{C}(0)=0.\nonumber
\end{align}

With the help of the unitary matrix $\mathbb{U}$ Eq.~\eqref{unitarym} we write the Hamiltonian Eq.~\eqref{schriffer} in diagonal form. The resulting Kondo Hamiltonian reads
 \begin{equation}\label{KondoH}
 \mathscr{H}_{\rm K}=\sum_a\Big(\mathscr{H}^a_0+\mathcal{J}_a\mathbf{s}_a\cdot\mathcal{S}\Big),
 \end{equation}
where $a={\rm e, o1, o2}$ are the channel indices and $\mathbf{s}_a$ stands for the spin-density operator in the new basis $a$. In addition, three nonzero eigenvalues (for $\varphi\neq 0$) of $\mathbb{J}$ have been relabeled as $\mathcal{J}_1=\mathcal{J}_{\rm e}$, $\mathcal{J}_2=\mathcal{J}_{\rm o1}$, and $\mathcal{J}_3=\mathcal{J}_{\rm o2}$. 

\textit{Kondo temperatures.} Going beyond the second order of the Schrieffer-Wolff transformation results in an interaction among difference channels which reads~\cite{twoc}
\begin{equation}\label{intt}
\mathscr{H}_{\rm ch-int}=-\sum_{a, b:a\neq b}\mathcal{J}_{ab} \mathbf{s}_a\cdot \mathbf{s}_b.
\end{equation}
While the amplitude of $\mathcal{J}_a$ (as seen earlier) scales with $\sim|t|^2/E_c$, the ferromagnetic coupling among different channels $\mathcal{J}_{ab}$ scales as $\sim\mathcal{J}_a\mathcal{J}_b/E_c$~\cite{twoc}. Therefore, Eq.~\eqref{intt} becomes irrelevant in the weak-coupling regime which allows us to define three distinct Kondo temperatures characterizing three different conduction channels,
\begin{equation}
\Tk^a=D\exp\left[-\frac{1}{2\nu_F \mathcal{J}_a}\right],
\end{equation}
where $D$ is a bandwidth and $\nu_F$ is the
three-dimensional electrons density of states in the leads. Since $\mathcal{J}_a(\varphi)$ are tuned through $\varphi$, we may express the Kondo temperature as $\Tk^a\equiv \Tk^a(\varphi)$, resulting in the least exotic situation $\Tk^{\rm e}>\Tk^{\rm o1}>\Tk^{\rm o2}$. We note that the last condition has been chosen just for the sake of simplicity, and arbitrary relations among Kondo temperatures can be considered straightforwardly with $\mathcal{J}_a$ presented earlier. As explained above, we consider the 3T Kondo setup in the presence of the Zeeman field $B$. The case with $B>\Tk^{\rm e}$ results in the weak-coupling regime of the problem. Decreasing $B$ below $\Tk^{\rm e}$ subsequently results in two intermediate states $\Tk^{\rm o2}<\Tk^{\rm o1}<B<\Tk^{\rm e}$ and $\Tk^{\rm o2}<B<\Tk^{\rm o1}<\Tk^{\rm e}$. Further decreasing of $B$ finally reaches the strong-coupling regime $B<\Tk^{\rm o2}$. In these ways, after three stages of screening of the $\mathcal{S}=3/2$ impurity spin by three conduction channels $\mathcal{K}=3$, a Kondo singlet is formed at the strong-coupling regime. While the weak-coupling and intermediate-coupling regimes can be understood in terms of well-known perturbative results~\cite{twoc}, the strong-coupling regime where all three Kondo channels are at resonance contains most of the nontrivial physics. Therefore, in the following, we develop the transport description at the strong-coupling regime of the 3SK impurity based on a local Fermi-liquid theory.

\textit{Scattering theory and conductance matrix.} We describe the transport at the strong-coupling regime of the 3SK effect by the celebrated Nozi\`{e}res FL theory~\cite{Nozieres} which allows us to express all of the scattering effect in terms of three-scattering phase shifts $\delta_{a\sigma}$ corresponding to three screening channels {\color{black} per spin projection $\sigma$}. The idea is to write the scattering matrix {\color{black} 
$\mathcal{A}_\sigma={\rm diag}\left\{e^{2 i \delta _{a\sigma}}\right\}$} for the three terminal geometry {\color{black} $a=1,2,3$} in the channel (rotated) diagonal basis as in Ref.~\cite{gpadd}.
%{\color{red}
%\begin{equation}
%\mathcal{A}_\sigma=\left(
%\begin{array}{ccc}
% e^{2 i \delta _{1\sigma}} & 0 & 0 \\
% 0 & e^{2 i \delta _{2\sigma}} & 0 \\
% 0 & 0 & e^{2 i \delta _{3\sigma}} \\
%\end{array}
%\right).
%\end{equation}
%}
From the unitary operator $\mathbb{U}(\varphi)$ Eq.~\eqref{unitarym} and {\color{black} $\mathcal{A}_\sigma$}, one then forms a scattering matrix characterizing the transport at the zero-temperature limit,
{\color{black}
\begin{equation}
\mathbb{S}(\varphi)\equiv\mathbb{U}^{\dagger}(\varphi)\mathcal{A}_\sigma\mathbb{U}(\varphi).
\end{equation}
}
The Landauer formula then expresses the conductance elements~\cite{land}
\begin{equation}\label{landauer}
G_{\rm \alpha\alpha'}(\varphi)=\frac{e^2}{h} \sum_{\sigma}\left|\mathbb{S}_{\rm \alpha\alpha'}(\varphi)\right|^2,
\end{equation}
where $e$ is the electron charge and $h$ is the Planck's constant. Equation.~\eqref{landauer} results in the conductance elements defined in the unit of $G_0=2e^2/h$ as
\begin{align}\label{conductances}
G_{12} &=\sum_\sigma\Big[4\mathcal{A}_1\sin ^2\delta _{12\sigma}{+}\frac{\mathcal{A}_2}{2} \sin ^2\delta _{23\sigma}-4\mathcal{A}_3\sin^2\delta_{13\sigma}\Big]\nonumber,\\
G_{13}& =G_{23}=8\mathcal{A}_3\sum_\sigma\sin^2\delta_{13\sigma},\;\;\delta_{\rm ij\sigma}\equiv\delta_{\rm i\sigma}-\delta_{\rm j\sigma}.
\end{align}
The other elements of the conductance matrix are expressible from the above presented elements by using the current conservation at each terminal and symmetry of the conductance elements. In the last equation, we defined the $\varphi$ dependent factors $\mathcal{A}_i(\varphi)$
\begin{equation}\nonumber
\mathcal{A}_1=\frac{1}{\mathcal{C}^2+4\mathcal{C}+12},\;\;\mathcal{A}_2=(\mathcal{C}+2)^2\mathcal{A}_1,\;\;\mathcal{A}_3=\mathcal{A}_1\mathcal{A}_2.
\end{equation}
If all three channels are at resonance, the Friedel sum rule guarantees that the
corresponding phase shift becomes $\delta_{\rm i\sigma}=\pi/2$. Equation.~\eqref{conductances} then accounts for the completely destructive interference among three resonance channels, thereby vanishing the conductance elements. At finite Zeeman field $B$, the phase shift deviates from the unitary limit.
The effects of finite $B\ll \Tk^{\rm o2}$ on the conductances are accounted for by the phase shift expansion based on the Nozi\`{e}res FL theory\cite{Nozieres, mora1, mora2, jvm},
\begin{equation}
\delta_{a\sigma}=\frac{\pi}{2}-\overline{\sigma}\alpha_a B,\;\;\alpha_a\simeq\frac{1}{\Tk^a}.
\end{equation}
The last equation provides the conductance element $G_{12}$ (in the unit of $G_0$) of the 3SK effect as
\begin{align}\label{ce}
G_{12}(\varphi, B) &=8\mathcal{A}_1\left(\frac{B}{\Tk^{\rm e}}-\frac{B}{\Tk^{\rm o1}}\right)^2+\mathcal{A}_2\left(\frac{B}{\Tk^{\rm o1}}-\frac{B}{\Tk^{\rm o2}}\right)^2\nonumber\\
&\;\;\;\;\;\;\;\;\;\;\;\;\;-8\mathcal{A}_3\left(\frac{B}{\Tk^{\rm e}}-\frac{B}{\Tk^{\rm o2}}\right)^2,
\end{align}
and similarly for other conductance and reflectance elements. All the features of two-stage and single-stage Kondo effects can be directly seen from Eq.~\eqref{ce}. Namely, for $\varphi=0$, the phase dependent parameter $\mathcal{C}(\varphi=0)=0$ and only the Kondo temperature of the even mode is nonzero $\Tk^{\rm e}\neq 0$. This describes the $\mathcal{S}=1/2$ Kondo impurity in 3T geometry with all the conductance elements being equal $G_{\alpha\alpha'}=4\left(B/\Tk^{\rm e}\right)^2/9$. Straightforward tuning of $\varphi$ also results in $\Tk^{\rm e, o1}\neq 0$ with $\Tk^{\rm o2}= 0$, which fully recovers the properties of two-stage Kondo effects.

The parallel configuration of higher-spin Kondo effects in multiterminal geometry results in just the additive contribution to the conductance $G_{\rm 3SK\; parallel}=6e^2/h$, and the corresponding series setups are very different with  $G_{\rm 3SK\; series}=\sum_{a\neq a'} \mathscr{L}_{aa'}\left[\alpha_a(\varphi)-\alpha_{a'}(\varphi)\right]^2B^2$. The factors $\mathscr{L}_{aa'}$ are tunable either by phase $\varphi$ or by tunneling asymmetry. We note that the naive expectation of conductance for 3SK effects, $G_{\rm 3SK}={\rm const}\sum_{a\neq a'} \left[\alpha_a-\alpha_{a'}\right]^2B^2$, is no longer correct, where the consistent description must find $\mathscr{L}_{aa'}(\varphi)$ and $\alpha_a(\varphi)$ carefully as we presented earlier. The calculation at finite temperature and voltage should be performed based on the low-energy FL Hamiltonian presented in SM~\cite{suppl} (which is left for future work).

The intermediate- and weak-coupling regimes of 3SK can be studied straightforwardly with the help of the above presented results and the well-known logarithmic decay of conductance at the weak-coupling regime~\cite{GP_Review_2005}. Namely, replacing the phase shifts appearing into our expression of conductance elements by~\cite{Coleman_PRB(75)_2007}
\begin{equation}
\delta_a(B)=\begin{cases}
\;\;\;\;\;\;{\rm const}\;\frac{\mathcal{S}}{\ln\left(B/\Tk^a\right)}, & B\gg \Tk^a\\
{\rm const}\left[1-\frac{\mathcal{S}}{\ln\left(\Tk^a/B\right)}\right], & B\ll \Tk^a,
\end{cases}
\end{equation}
provides full access to uncover the transport descriptions at the intermediate- and weak-coupling regimes. All the features of 2SK \DK{and 1SK} effects are thus captured by the 3SK model presented here, in addition to providing different insights on the Kondo paradigm associated with high-spin states. Therefore, the independent control of interfering Kondo channels and their interplay with each other might provide an effective way of using the 3SK setup as a quantum interferometer (further details have been presented in SM).

\textit{Summary.} 
We presented a simple description of the multistage Kondo effect in multiterminal geometry based on the Nozi\`{e}res Fermi-liquid theory. The studied \AP{framework} describes intrinsically multiterminal effects and allows for a
precise discrimination between different configurations of the electron states. This provides an access to very
rich physics beyond the commonly studied two-terminal and one- or two-mode Kondo
screening. We uncovered various, albeit simple, ways of fine-tuning the multiresonant Kondo channels
and their interplay with each other in order to observe the constructive/destructive interference in the
simplest possible setup. This minimal setup of the three-stage Kondo effect can be
used as a quantum interferometer which also contains all the physics associated with the two-terminal Kondo
paradigm and at the same time allows a straightforward generalization to other numbers of stages. The developed framework provides a controllable way to construct a desired realization of the
Kondo effect with a particular number of stages, terminals, and channels from combinations of elementary ``building blocks," conceptually alike to making complicated constructions from simple blocks (consequently referred to as a modular quantum interferometer - see Fig. \ref{cond_1}). The studied transport observables are within the reach of existing experimental setups, such as of the recent experiment on 2SK effects~\cite{chines_2sk}. Therefore, we believe that the presented \AP{ideas} would motivate further experiments as well as theoretical works to uncover the Kondo paradigm with high-spin states.

\textit{Acknowledgments.} We are thankful to Jan von Delft and Seung-Sup Lee for inspiring discussions. The work
of M.K. is conducted within the framework of the Trieste Institute
for Theoretical Quantum Technologies (TQT). M.K. appreciates the hospitality of the Physics Department,
Arnold Sommerfeld Center for Theoretical Physics and Center for NanoScience, Ludwig-Maximilians-Universit{\"a}t M{\"u}nchen, where part of this work has been performed.
%\bibliography{3S3T}
\begin{widetext}
\section*{Supplemental material}

In this Supplemental Materials we present additional details for the derivation of key equations. All used notations are in accordance with the main text.

%%%%%%%%%%%%%%%%%%%%%%%%%%%%%%%%%%%%%%% TITLE %%%%%%%%%%%%%%%%%%%%%%%%%%%%%%%%%%%%

\subsection{Three-terminal setup}

We consider a three level ($i=1, 2, 3$) quantum impurity tunnel-coupled to three non-interacting reservoirs ($\alpha=L,M,R$ - Left, Middle and Right). The impurity is described by three-orbital Anderson model (as presented in the main text). The Schrieffer-Wolff transformation of Anderson model eliminates the charge fluctuations among the orbitals resulting in the effective Hamiltonian
\begin{equation} H_{eff}=\sum_{k}\sum_{\alpha=L,M,R} \sum_{\sigma=\uparrow,\downarrow}\epsilon_{k}C^{\dagger}_{\alpha k\sigma}C_{\alpha k \sigma}+\sum_{\alpha,\alpha^{\prime}=L,M,R}\mathcal{J}_{\alpha\alpha^{\prime}}\mathbf{s}_{\alpha\alpha^{\prime}}\mathbf{S}, \end{equation}
where the matrix of exchange coupling is represented by $\mathcal{J}_{\alpha,\alpha^{\prime}}$ and the spin density of conduction electron writes
\begin{equation} \mathbf{s}_{\alpha\alpha^{\prime}}=\frac{1}{2}\sum_{\sigma,\sigma^{\prime}}C^{\dagger}_{\alpha k\sigma}\mathbf{\tau}_{\sigma\sigma^{\prime}}C_{\alpha k \sigma^{\prime}}. \end{equation}
Denoting the tunneling elements from the lead $\alpha$ to the $i$-th orbital by $t_{\alpha i}$, we obtain the exchange matrix
\begin{equation} \label{Jttt} J_{\alpha\alpha^{\prime}}=\frac{2}{SE_C}\left(\begin{matrix}
|t_{L1}|^2+|t_{L2}|^2+|t_{L3}|^2 & t_{L1}^*t_{M1}+t_{L2}^*t_{M2}+t_{L3}^*t_{M3} & t_{L1}^*t_{R1}+t_{L2}^*t_{R2}+t_{L3}^*t_{R3}\\
t_{M1}^*t_{L1}+t_{M2}^*t_{L2}+t_{M3}^*t_{L3} & |t_{M1}|^2+|t_{M2}|^2+|t_{M3}|^2 & t_{M1}^*t_{R1}+t_{M2}^*t_{R2}+t_{M3}^*t_{R3}\\
t_{R1}^*t_{L1}+t_{R2}^*t_{L2}+t_{R3}^*t_{L3}& t_{R1}^*t_{M1}+t_{R2}^*t_{M2}+t_{R3}^*t_{M3} & |t_{R1}|^2+|t_{R2}|^2+|t_{R3}|^2
\end{matrix} \right).\end{equation}
 
Let us denote the eigenvalues of this matrix as $\mathcal{J}_1$, $\mathcal{J}_2$, $\mathcal{J}_3$.\\ 
Any $3\times3$ square matrix has three invariants, they are
\begin{equation} \label{trJ} Tr(\mathbb{J})=\mathcal{J}_1+\mathcal{J}_2+\mathcal{J}_3=\frac{2}{\mathcal{S}E_c}\sum_{\alpha={L, M, R}}\sum_{i=1}^3|t_{\alpha i}|^2, \end{equation}
\begin{equation} \label{trJtr} \frac{1}{2}\left[\left(Tr(\mathbb{J})\right)^2-Tr(\mathbb{J}^2) \right] =\mathcal{J}_1\mathcal{J}_2+\mathcal{J}_2\mathcal{J}_3+\mathcal{J}_1\mathcal{J}_3=\left(\frac{2}{\mathcal{S}E_c}\right)^2\sum_{\alpha\beta}\sum_{i\neq j}|t_{\alpha i}t_{\beta j}-t_{\alpha j}t_{\beta i}|^2, \end{equation}
\begin{equation} \label{detJ} Det(\mathbb{J})=\mathcal{J}_1\mathcal{J}_2\mathcal{J}_3=\left(\frac{2}{\mathcal{S}E_c}\right)^3\left|\sum_{ijk=1}^3\varepsilon^{ijk}t_{Li}t_{Mj}t_{Rk}\right |^2, \end{equation}
$\varepsilon^{ijk}$ is the antisymmetric Levi-Civita tensor. 
Eqs. (\ref{trJ}-\ref{detJ}) are related to the Vieta's formulas for a cubic polynomial. These formulas state that roots $r_1, r_2, r_3$ of a cubic polynomial $P(\lambda)=\lambda^3+a\lambda^2+b\lambda+c$ satisfy
\begin{align} \nonumber r_1+r_2+r_3=-a,\\
\nonumber r_1r_2+r_1r_3+r_2r_3=b,\\
\nonumber r_1r_2r_3=-c.
\end{align}
The polynomial $P(\lambda)$ corresponds to the equation for eigenvalues of Eq. (\ref{Jttt}), 
\begin{align*}
\lambda^3-\lambda^2 \frac{2}{\mathcal{S}E_c}\sum_{\alpha, i}|t_{\alpha i}|^2+\lambda \left(\frac{2}{\mathcal{S}E_c}\right)^2\sum_{\alpha\beta}\sum_{i\neq j}|t_{\alpha i}t_{\beta j}-t_{\alpha_j}t_{\beta i}|^2- \left(\frac{2}{\mathcal{S}E_c}\right)^3\left|\sum_{ijk}\varepsilon^{ijk}t_{Li}t_{Mj}t_{Rk}| \right|^2=0.
\end{align*}

Let us find conditions allowing Eq.(\ref{detJ}) to be positive. It allows the matrix Eq.(\ref{Jttt}) to have three non-zero eigenvalues ($\mathcal{J}_1\mathcal{J}_2\mathcal{J}_3\neq 0$).\\
We choose the tunneling coefficients $t_{\alpha i}$ as 
\begin{align}
 \nonumber t_{L 1}=|t|e^{\textit{i}\varphi_1},\, t_{L 2}=|t|e^{\textit{i}\varphi_2},\, t_{L 3}=|t|e^{\textit{i}\varphi_3}, \\ \label{TunnelingElements} t_{M 1}=|t|e^{\textit{i}\varphi_4},\, t_{M 2}=|t|e^{\textit{i}\varphi_5},\, t_{M 3}=|t|e^{\textit{i}\varphi_6},\\
  \nonumber t_{R 1}=t_{R 2}=t_{R 3}=|t|. \end{align}
Eq.(\ref{detJ}) is now equivalent to
\begin{equation} \label{detCondition} Det(\mathbb{J})\sim |t|^2\left | e^{\textit{i}(\varphi_1+\varphi_5)}-e^{\textit{i}(\varphi_1+\varphi_6)}+e^{\textit{i}(\varphi_3+\varphi_4)}-e^{\textit{i}(\varphi_3+\varphi_5)}+e^{\textit{i}(\varphi_2+\varphi_6)}-e^{\textit{i}(\varphi_2+\varphi_4)} \right |^2. \end{equation}

To satisfy $Det(\mathbb{J})>0$, it is enough to choose two phases from different terminals with different $i, j$ indexes as non-zero (e.g. $\varphi_1\neq 0$ and $\varphi_5\neq 0$, or other combinations corresponding to sums in the exponents above) with all other phases being zero. Naturally, this condition guaranties that (\ref{trJ}) and (\ref{trJtr}) are also positive. It is easy to see that if all the phases related to either $L$ or $M$ terminal are equal (e.g. $\varphi_1=\varphi_2=\varphi_3$), the determinant (\ref{detCondition}) is zero. If we choose $\varphi_1\neq 0$, $\varphi_4\neq 0$ and all other $\varphi_i=0$, we again obtain $Det(\mathbb{J})=0$.\\

Let us check whether it's possible to have three identical non-zero eigenvalues (i.e. three identical finite Kondo temperatures). This condition is satisfied if all non-diagonal elements of matrix (\ref{Jttt}) are zero. Choosing the tunneling elements in the form of Eq.(\ref{TunnelingElements}), we get the following conditions
\begin{align*}
e^{\textit{i}(\varphi_4-\varphi_1)}+e^{\textit{i}(\varphi_5-\varphi_2)}+e^{\textit{i}(\varphi_6-\varphi_3)}=0,\\
e^{-\textit{i}\varphi_1}+e^{-\textit{i}\varphi_2}+e^{-\textit{i}\varphi_3}=0,\\
e^{-\textit{i}\varphi_4}+e^{-\textit{i}\varphi_5}+e^{-\textit{i}\varphi_6}=0.
\end{align*}
These equations are satisfied for $\varphi_2=\varphi_5=0$, $\varphi_1=\varphi_6=-\varphi_3=-\varphi_4=\{\frac{2\pi}{3}, \, \frac{4\pi}{3}\}$, and we get $\mathcal{J}_1=\mathcal{J}_2=\mathcal{J}_3=3\mathcal{J}_0$ with $\mathcal{J}_0=\frac{2 |t|^2}{SE_C}$, so it's possible to have three identical Kondo temperatures.\\
Let us choose 
\begin{align}
\nonumber t_{L1}=|t|e^{\textit{i}\varphi},\, t_{L2}=|t|,\, t_{L3}=|t|e^{-\textit{i}\varphi},\\
\label{TunnelingElements_new} t_{M1}=|t|e^{-\textit{i}\varphi},\, t_{M2}=|t|,\, t_{M3}=|t|e^{\textit{i}\varphi},\\
\nonumber t_{R1}=t_{R2}=t_{R3},
\end{align}
with arbitrary phase $\varphi$. Manipulating this phase, one can have any possible relation between Kondo temperatures. The matrix $\mathbb{J}$ takes form
\begin{align} \label{JmatrixT}
\mathbb{J} = \mathcal{J}_0 \begin{pmatrix}
3 & 1+2\cos 2\phi & 1+2\cos\varphi\\
1+2\cos 2\varphi & 3 & 1+2\cos\varphi \\
1+2\cos\varphi & 1+2\cos\varphi & 3
\end{pmatrix} ,
\end{align}
here we chose $t_{R1}=t_{R2}=t_{R3}=|t|$ (other choices that do not affect eigenvalues of this matrix are possible, see discussion below).\\

The eigenvalues of this matrix are
\begin{align} \label{Jphase}
\mathcal{J}_{1,3}=\frac{\mathcal{J}_0}{2}\left(7+2\cos2\varphi\pm \sqrt{32\cos\varphi+(5+2\cos2\varphi)^2} \right),\,\,\,\mathcal{J}_2=4\mathcal{J}_0\sin^2\varphi.
\end{align}
In general, these eigenvalues give us three different corresponding non-zero Kondo temperatures, but for specific values of $\varphi$ we can have $\mathcal{J}_2=\mathcal{J}_3=0,\, \mathcal{J}_1>0$ ($\varphi=0$); $\mathcal{J}_3=0$, $\mathcal{J}_{1,2}>0$ ($\varphi=\pi$); $\mathcal{J}_1=\mathcal{J}_3>\mathcal{J}_2>0$ ($\varphi=\frac{\pi}{2}$); $\mathcal{J}_1=\mathcal{J}_2=\mathcal{J}_3>0$ ($\varphi=\frac{2\pi}{3},\, \frac{4\pi}{3}$). Eigenvalues $\mathcal{J}_1,\, \mathcal{J}_2,\, \mathcal{J}_3$ as functions of $\varphi$ are given in Fig.\ref{fig:Jvalues} of the main text. Redefining eigenvalue indexes $\{1,2,3\}\rightarrow \{e,o1,o2\}$ to have a hierarchy $\mathcal{J}_e>\mathcal{J}_{o1}>\mathcal{J}_{o2}$, we have the corresponding Kondo temperatures $T_{K}^{a}\sim  D \exp\left(-\frac{1}{2\nu_F \mathcal{J}_{a}(\varphi)}\right)$, $a=\{e,o1,o2\}$, $D$ is a bandwidth of conduction electrons band, $\nu_F$ is the density of states. Note that the choice of the phase for the tunneling constants associated with the third terminal (R) in Eq.(\ref{TunnelingElements_new}) does not affect the eigenvalues as long that these tunneling constants have the same phase, since that does not change Eq.(\ref{detJ}) (we can choose $t_{R1}=t_{R2}=t_{R3}=|t|$ for simplicity).\\

\section{energy gap}
Now we write down the wave functions of the electrons in the dot. The dot under consideration has spin $S=3/2$. The three electrons form 8 states ($2\times 2\times 2$): one quartet with total spin $3/2$ and two doublets with total state $1/2$ each.
\begin{equation}
\nonumber \left|\frac{3}{2},\frac{3}{2} \right\rangle=\left|\uparrow\uparrow\uparrow \right\rangle,
\end{equation}
\begin{equation}
\nonumber \left|\frac{3}{2},\frac{1}{2} \right\rangle=\frac{1}{\sqrt{3}}\left(\left|\downarrow\uparrow\uparrow \right\rangle+\left|\uparrow\downarrow\uparrow \right\rangle+\left|\uparrow\uparrow\downarrow \right\rangle\right),
\end{equation}
\begin{equation}
\nonumber \left|\frac{3}{2},-\frac{1}{2} \right\rangle=\frac{1}{\sqrt{3}}\left(\left|\downarrow\downarrow\uparrow \right\rangle+\left|\uparrow\downarrow\downarrow \right\rangle+\left|\downarrow\uparrow\downarrow \right\rangle\right),
\end{equation}
\begin{equation}
\nonumber \left|\frac{3}{2},-\frac{3}{2} \right\rangle=\left|\downarrow\downarrow\downarrow \right\rangle,
\end{equation}
\begin{equation}
\nonumber \left|\frac{1}{2},\frac{1}{2} \right\rangle_1=\frac{1}{\sqrt{2}}\left(\left|\downarrow\uparrow\uparrow \right\rangle-\left|\uparrow\uparrow\downarrow \right\rangle\right),
\end{equation}
\begin{equation}
\nonumber \left|\frac{1}{2},-\frac{1}{2} \right\rangle_1=\frac{1}{\sqrt{2}}\left(\left|\downarrow\downarrow\uparrow \right\rangle-\left|\uparrow\downarrow\downarrow \right\rangle\right),
\end{equation}
\begin{equation}
\nonumber \left|\frac{1}{2},\frac{1}{2} \right\rangle_2=\frac{1}{\sqrt{6}}\left(\left|\downarrow\uparrow\uparrow \right\rangle-2\left|\uparrow\downarrow\uparrow \right\rangle +\left|\uparrow\uparrow\downarrow \right\rangle\right),
\end{equation}
\begin{equation}
\nonumber \left|\frac{1}{2},-\frac{1}{2} \right\rangle_2=\frac{1}{\sqrt{6}}\left(2\left|\downarrow\uparrow\downarrow \right\rangle-\left|\downarrow\downarrow\uparrow \right\rangle -\left|\uparrow\downarrow\downarrow \right\rangle\right).
\end{equation}

Let us find the ground state configuration of the dot. For that, we suppose that we have three spins $S_1$, $S_2$, $S_3$ on a ring interacting via the ferromagnetic Heisenberg Hamiltonian. This interaction allows us to reproduce the Hund's rules in the dot.
\begin{align*}
H=\mathcal{I}(\mathbf{S}_1\mathbf{S}_2+\mathbf{S}_2\mathbf{S}_3+\mathbf{S}_1\mathbf{S}_3),\,\,\, (\mathcal{I}<0).
\end{align*}
The scalar product in the brackets reads
\begin{align*}
(\mathbf{S}_1\mathbf{S}_2+\mathbf{S}_2\mathbf{S}_3+\mathbf{S}_1\mathbf{S}_3)=\frac{1}{2}\left(\mathbf{S}_1+\mathbf{S}_2+\mathbf{S}_3 \right)^2-\frac{1}{2}\left(\mathbf{S}^2_1+\mathbf{S}_2^2+\mathbf{S}_3^2\right).
\end{align*}
The spin $3/2$ (quartet) state and the $1/2$ (doublet) states are separated by the gap
\begin{align} \label{Hund}
\Delta=E(S_{total}=1/2)-E(S_{total}=3/2)=-\frac{|\mathcal{I}|}{2}\left(\frac{1}{2}\cdot\frac{3}{2}-3\cdot\frac{3}{4}\right)-\frac{|\mathcal{I}|}{2}\left(\frac{3}{2}\cdot\frac{5}{2}-3\cdot\frac{3}{4} \right)=\frac{3}{2}|\mathcal{I}|,
%E(S_{total}=1/2)=-\frac{|\mathcal{I}|}{2}\left(\frac{1}{2}\cdot
%\frac{3}{2}-3\cdot\frac{3}{4}\right)=\frac{3}{4}|\mathcal{I}|,
\end{align}
so the ground state is the quartet, the doublets are excited states separated from it by the gap $\Delta$. This quartet, appearing as the ground state from Eq.(\ref{Hund}) corresponds to the ground state of dot considered in the main text. \\
\section{Rotation of electron states}
Now we discuss a generalization of the Glazman-Raikh (GR) rotation on the three-terminal case. Let us construct an effective basis with one even and two odd states $C_e,\, C_{o1},\, C_{o2}$ out of three original states of the leads $C_L,\, C_M,\, C_R$. The original GR approach deals with the two-channel case, so the transformation is given by the real-valued matrix $\mathbb{U}_2$ of the $SU(2)$ representation (the rotation matrix in 2D) defined by one parameter $\alpha$:
\begin{align} \label{2dmatrix}
\begin{pmatrix}
C_e\\
C_o
\end{pmatrix}=\begin{pmatrix}
\cos \alpha & \sin \alpha\\
-\sin \alpha & \cos \alpha
\end{pmatrix} \begin{pmatrix}
C_L\\ 
C_R
\end{pmatrix}.
\end{align}
Let us start with a two-terminal two-channels case and find $\alpha$ for this case. There we have 4 tunneling parameters chosen to be real for simplicity. $\mathbb{J}_{2\times 2}$ matrix composed of $\mathcal{J}_{\alpha\alpha^{\prime}}$ tunneling elements for the two-terminal two-channel case is 
\begin{align*}
\mathbb{J}_{2\times 2}=\frac{2}{\mathcal{S}E_c}\begin{pmatrix}
t^2_{L1}+t^2_{L2} & t_{L1}t_{R1}+t_{L2}t_{R2}\\
t_{L1}t_{R1}+t_{L2}t_{R2} & t^2_{R1}+t^2_{R2}
\end{pmatrix},
\end{align*} and we parametrize tunneling elements as
\begin{align*}
t_{L1}=t\cos\theta\cos\varphi_L;
t_{L2}=t\cos\theta\sin\varphi_L;
t_{R1}=t\sin\theta\cos\varphi_R;
t_{R1}=t\sin\theta\cos\varphi_R.
\end{align*}
The transformation $\mathbb{U}_2\mathbb{J}_{2\times 2}\mathbb{U}^{-1}_2$ ($\mathbb{U}_2$ is the rotation matrix from Eq.(\ref{2dmatrix})) diagonalizes the $\mathbb{J}_{2\times 2}$ matrix, this condition gives us the relation between angle $\alpha$ and angles $\theta$, $\varphi_L$, $\varphi_R$:
\begin{align*}
\tan(2\alpha)=\tan(2\theta)\cos(\varphi_L-\varphi_R),
\end{align*}
so $\alpha=\theta$ when all the tunneling amplitudes are equal.\\
In addition, we have the following relation for the eigenvalues $\tilde{\mathcal{J}}_1$ and $\tilde{\mathcal{J}}_2$ of the $\mathbb{J}_{2\times 2}$ matrix
\begin{align*}
\tilde{\mathcal{J}}_1+\tilde{\mathcal{J}}_2= \mathcal{J}_0,\\
\tilde{\mathcal{J}}_1\tilde{\mathcal{J}}_2=\frac{\mathcal{J}_0^2}{4}\sin^2(2\theta)\sin^2(\varphi_L-\varphi_R).
\end{align*}
 where $\mathcal{J}_0=2 t^2/E_c$ for $\mathcal{S}=1$.

A general real-valued matrix of the $SU(3)$ representation is characterized by three angles $\alpha,\, \beta,\, \gamma$, this matrix is composed of the eigenvector for the symmetric mode and modes orthogonal to it~\cite{lastref1, lastref2, apadd}:
\begin{align} \label{Rotation.general} \begin{pmatrix}
C_e\\
C_{o1}\\
C_{o2}
\end{pmatrix}=
\begin{pmatrix}
\cos \alpha \cos \beta & \sin \alpha & \cos\alpha\sin\beta \\
\sin\beta\sin\gamma-\sin\alpha\cos\beta\cos\gamma & \cos\alpha\cos\gamma & -\cos\beta\sin\gamma-\sin\alpha\sin\beta\cos\gamma \\
-\sin\alpha\cos\beta\sin\gamma-\sin\beta\cos\gamma & \cos\alpha\sin\gamma & \cos\beta\cos\gamma-\sin\alpha\sin\beta\sin\gamma
\end{pmatrix} \begin{pmatrix}
C_L\\ 
C_M\\
C_R
\end{pmatrix}.
\end{align}
$\gamma$ defines a rotation of two mutually orthogonal "odd" states in a plane perpendicular to the even state (any two mutually orthogonal vectors belonging to the plane can be chosen as $C_{o1}$ and $C_{o2}$). %Tunneling coefficients can be expressed via $\alpha$ and $\beta$ (for instance, $\frac{|t_M|}{|t_R|}=\tan\theta_1\csc\theta_2$, $\frac{|t_M|}{|t_L|}=\tan\theta_1\sec\beta$).
 We choose $\gamma=\frac{\pi}{2}$ for simplicity (so the $C_{o1}$ and $C_{o2}$ states become antisymmetric), the generalized GR transformation then reads
\begin{align}  \label{Matrix.rotation} \begin{pmatrix}
C_e\\
C_{o1}\\
C_{o2}
\end{pmatrix}=
\begin{pmatrix}
\cos \alpha \cos \beta & \sin \alpha & \cos\alpha\sin\beta \\
\sin\beta & 0 & -\cos\beta \\
-\sin\alpha\cos\beta & \cos\alpha & -\sin\alpha\sin\beta
\end{pmatrix} \begin{pmatrix}
C_L\\ 
C_M\\
C_R
\end{pmatrix}.
\end{align}
Other choices of $\gamma$ are possible. For instance, putting $\gamma=0$, we obtain the same matrix of Eq.(\ref{Matrix.rotation}) with flipped second and third lines (the new third line additionally changes signs). In a general case of the arbitrary $\gamma$ angle, we get a matrix with its lines being linear combinations of the lines (\ref{Matrix.rotation}).\\
In general, the angles $\alpha$, $\beta$, $\gamma$ are expressed via the Euler angles and incorporate the asymmetries between the channels, the matrix in Eq.(\ref{Rotation.general}) is a rotation matrix. It can be obtained by the rotation $R_z(\gamma)R_y(-\alpha)R_x(\beta)$, up to cyclic permutations, which mean simply relabelling of the axes. $R_i$ are the rotation operators around the corresponding axis $i$ on the angle given angle. 

For $|t_L|=|t_M|=|t_R|$, we have $\alpha=\arcsin\frac{1}{\sqrt{3}}$, $\beta=\frac{\pi}{4}$, so the transformation becomes
\begin{align*}\begin{pmatrix}
C_e\\
C_{o1}\\
C_{o2}
\end{pmatrix}=
\begin{pmatrix}
\frac{1}{\sqrt{3}} & \frac{1}{\sqrt{3}} & \frac{1}{\sqrt{3}} \\
\frac{1}{\sqrt{2}} & 0 & -\frac{1}{\sqrt{2}} \\
-\frac{1}{\sqrt{6}} & \frac{2}{\sqrt{6}} & -\frac{1}{\sqrt{6}}
\end{pmatrix} \begin{pmatrix}
c_L\\ 
c_M\\
c_R
\end{pmatrix}.
\end{align*}

The three terminals $L,\, M,\, R$ are equivalent, but the parametrization (\ref{Rotation.general}) breaks the symmetry between them, so the cyclic permutations of the leads are possible: $L\rightarrow M\rightarrow R$ ~\cite{mk}. For instance, the permutation $\{R\rightarrow L,\, L\rightarrow M,\,M\rightarrow R\}$ gives
\begin{align*}\begin{pmatrix}
C_e \\
\tilde{C}_{o1}\\
\tilde{C}_{o2}
\end{pmatrix}=
\begin{pmatrix}
\frac{1}{\sqrt{3}} & \frac{1}{\sqrt{3}} & \frac{1}{\sqrt{3}} \\
-\frac{1}{\sqrt{2}} &  \frac{1}{\sqrt{2}} & 0 \\
-\frac{1}{\sqrt{6}}& -\frac{1}{\sqrt{6}} & \frac{2}{\sqrt{6}}
\end{pmatrix} \begin{pmatrix}
C_L\\ 
C_M\\
C_R
\end{pmatrix}.
\end{align*}
We will use this representation further throughout the paper. The observables do not depend on our choice of parametrization, so they must be averaged over the $\gamma$-angle.
\\
\subsection{Phase tunable regime}
Now, let us consider a general case when the tunneling coefficients are complex and depend of the Aharonov-Bohm phases (\ref{TunnelingElements_new}). The eigenvectors corresponding to the three non-zero eigenvalues (\ref{Jphase}) cannot be chosen to be the phase-independent in the three-terminal setup. The corresponding matrix $\mathbb{U}$ that diagonalizes Eq.(\ref{Jttt}) is composed of the eigenvectors of $\mathbb{J}$ and reads

\begin{align*}
\mathbb{U}(\varphi)=\begin{pmatrix}
v_{11}(\varphi) & v_{12}(\varphi) & v_{13}(\varphi)\\
-\frac{1}{\sqrt{2}} & \frac{1}{\sqrt{2}} & 0\\
v_{31}(\varphi) & v_{32}(\varphi) & v_{33}(\varphi)
\end{pmatrix},
\end{align*}
where 
\begin{align*}
v_{11}(\varphi)=v_{12}(\varphi)\equiv \frac{-1+2\cos\varphi+\sqrt{11-4\cos\varphi+2\cos2\varphi}}{4\sqrt{1+\frac{1}{8}\left(-1+2\cos\varphi+\sqrt{11-4\cos\varphi+2\cos2\varphi}\right)^2}},\\
v_{13}(\varphi)\equiv \frac{1}{\sqrt{1+\frac{1}{8}\left(-1+2\cos\varphi+\sqrt{11-4\cos\varphi+2\cos2\varphi}\right)^2}},\\
v_{31}(\varphi)=v_{32}(\varphi)\equiv \frac{-1+2\cos\varphi-\sqrt{11-4\cos\varphi+2\cos2\varphi}}{4\sqrt{1+\frac{1}{8}\left(-1+2\cos\varphi-\sqrt{11-4\cos\varphi+2\cos2\varphi}\right)^2}},\\
v_{33}(\varphi)\equiv  \frac{1}{\sqrt{1+\frac{1}{8}\left(-1+2\cos\varphi-\sqrt{11-4\cos\varphi+2\cos2\varphi}\right)^2}}.
\end{align*}
This matrix is identical to Eq.(\ref{unitarym}) of the main text.  

Note that due to presence of the phases in particular tunneling elements, the symmetry between the terminals is broken. In particular, choice (\ref{TunnelingElements_new}) makes terminal R different from terminals L and M.

\subsection{Tunneling amplitudes tunable regime}
\setcounter{equation}{3}
To ensure that Eq.(\ref{detJ}) is positive (i.e. there are three non-zero Kondo temperatures), one does not have to alter phases of the tunneling elements, since the same effect can be achieved by changing amplitudes of these elements. In general, tunneling processes of $N$-terminal $M$-channel Kondo problem are parametrized by $2\times N\times  M$ real parameters. The condition for $N$ nonzero Kondo temperatures ($det(\mathbb{J})\neq 0$) puts a constrain on a minimal number of parameters that must be tuned (i.e. a certain number of symmetries must be broken). On the other hand, if one wants to have a system where the Kondo effect splits into symmetric and antisymmetric channels, one has to impose a number of constrains that reduce the number of adjustable parameters. Effectively, there is a $K$-dimensional surface in the $2\times N\times M$-dimensional parametric space ($0<K<2\times N\times M$) which satisfies the necessary conditions. \\
Let us illustrate it for the 3-terminal 3-channel Kondo problem under consideration. There are 18 parameters which we reparametrize as
\begin{align} \label{amplitudes}
\nonumber t_{L1}=t\sin\theta_1\sin\theta_2\sin\phi_{La}\sin\phi_{Lb}e^{\textit{i}\psi_{L1}}, \\
\nonumber t_{L2}=t\sin\theta_1\sin\theta_2\cos\phi_{La}e^{\textit{i}\psi_{L2}}, \\
\nonumber t_{L3}=t\sin\theta_1\sin\theta_2\sin\phi_{La}\cos\phi_{Lb}e^{\textit{i} \psi_{L3}},\\
\nonumber t_{M1}=t\cos\theta_1\sin\phi_{Ma}\sin\phi_{Mb}e^{\textit{i}\psi_{M1}}, \\
t_{M2}=t\cos\theta_1\cos\phi_{Ma}e^{\textit{i}\psi_{M2}},\\
\nonumber t_{M3}=t\cos\theta_1\sin\phi_{La}\cos\phi_{Mb}e^{\textit{i}\psi_{M3}}, \\
\nonumber t_{R1}=t\sin\theta_1\cos\theta_2\sin\phi_{Ra}\sin\phi_{Rb}e^{\textit{i}\psi_{R1}}, \\
\nonumber t_{R2}=t\sin\theta_1\cos\theta_2\cos\phi_{Ra}e^{\textit{i}\psi_{R2}}, \\
\nonumber t_{R3}=t\sin\theta_1\cos\theta_2\sin\phi_{Ra}\cos\phi_{Rb}e^{\textit{i}\psi_{R3}},
\end{align}
\begin{align*}
\sin\phi_{Lb}=\frac{|t_{L1}|}{\sqrt{|t^2_{L1}|+|t^2_{L3}|}},\,\,\,\sin\phi_{Mb}=\frac{|t_{M1}|}{\sqrt{|t^2_{M1}|+|t^2_{M3}|}},\,\,\,
\sin\phi_{Rb}=\frac{|t_{R1}|}{\sqrt{|t^2_{R1}|+|t^2_{R3}|}}, \\
t_L=\sqrt{|t^2_{L1}|+|t^2_{L2}|+|t^2_{L3}|},\,\,\, t_M=\sqrt{|t^2_{M1}|+|t^2_{M2}|+|t^2_{M3}|},\,\,\, t_R=\sqrt{|t^2_{R1}|+|t^2_{R2}|+|t^2_{R3}|}, \\
\sin\phi_{La}=\frac{\sqrt{|t^2_{L1}|+|t^2_{L3}|}}{t_L}, \,\,\, \sin\phi_{Ma}=\frac{\sqrt{|t^2_{M1}|+|t^2_{M3}|}}{t_M}, \,\,\, \sin\phi_{Ra}=\frac{\sqrt{|t^2_{R1}|+|t^2_{R3}|}}{t_R}, \\
\sin\theta_2=\frac{t_L}{\sqrt{t_L^2+t_R^2}} ,\,\,\, t=\sqrt{t_L^2+t_M^2+t_R^2},\,\,\,\sin\theta_1=\frac{\sqrt{t_L^2+t_R^2}}{t},\\
\theta_1\, \theta_2,\, \phi_{La}, \, \phi_{Lb},\, \phi_{Ma},\, \phi_{Mb},\, \phi_{Ra},\, \phi_{Rb} \in [0,\frac{\pi}{2}].
\end{align*}
Parameter $t$ defines the absolute values of the Kondo temperatures but does not affect the rotation matrix $\mathbb{U}$.
Let us choose all tunneling phases $\psi_{\alpha,i}$ to be zero. The most symmetric case (with all terminals and all channels being identical) corresponds to values $\theta_1=\arccos\frac{1}{\sqrt{3}}$, $\theta_2=\frac{\pi}{4}$, $\phi_{\alpha a}=\arccos\frac{1}{\sqrt{3}}$, $\phi_{\alpha b}=\frac{\pi}{4}$, $\alpha=\{L,M,R\}$. Now we break the symmetry between different channels in two terminals and introduce an asymmetry parameter $\varphi$ so that
\begin{align*}
\phi_{Lb}=\frac{\pi}{4}+\varphi,\, \phi_{Mb}=\frac{\pi}{4}-\varphi,\, \phi_{Rb}=\frac{\pi}{4}.
\end{align*}
Plugging tunneling elements (\ref{amplitudes}) into Eq.(\ref{Jttt}), we exactly reproduce the $\mathbb{J}$ matrix (\ref{JmatrixT}) and all the further calculations become identical to the case we considered above, where instead of asymmetry in tunneling amplitudes, asymmetry in tunneling phases was introduced.\\
We have here 
\begin{align*}
\mathcal{J}_1+\mathcal{J}_2+\mathcal{J}_3&=&9 \mathcal{J}_0;\\
\mathcal{J}_1\mathcal{J}_2+\mathcal{J}_1\mathcal{J}_3+\mathcal{J}_2\mathcal{J}_3&=&2\mathcal{J}_0^2\left(9-4\cos\varphi-4\cos2\varphi-\cos4\varphi\right);\\
\mathcal{J}_1\mathcal{J}_2\mathcal{J}_3&=&64  \mathcal{J}_0^3\sin^4\left(\frac{\varphi}{2}\right)\sin^2\varphi.
\end{align*}
Note that we define here $\mathcal{J}_0=2|t|^2/({\mathcal S} E_c)=2t^2/(9{\mathcal S} E_c)$.
Also angles $\alpha$, $\beta$ and $\gamma$ acquire dependence on $\varphi$.

\section{Strong-coupling fixed point}
The strong coupling fixed point Hamiltonian contains six leading irrelevant operators \cite{twoc}:
$$
H=-\sum_{i=1}^3\lambda_i :\vec{s}_i(0)\cdot\vec{s}_i(0): - \sum_{i=1}^3\sum_{j\neq i}^3 \lambda_{ij} :\vec{s}_i(0)\cdot\vec{s}_j(0):
$$
There are two important limiting cases. The first case corresponds to the situation when all three eigenvalues of the exchange matrix (and therefore all three Kondo temperatures) are equal.
\begin{itemize}
\item $\lambda_1=\lambda_2=\lambda_3=\lambda_{12}=\lambda_{23}=\lambda_{13}=\Lambda$
$$H= -\Lambda :\left(\vec{s}_1(0)+\vec{s}_2(0)+\vec{s}_3(0)\right)^2:$$
\end{itemize}
This is the case of total destructive interference. The net current through the system is zero.

The second important limiting case corresponds to degeneracy of two eigenvalues corresponding to two
orthogonal anti-symmetric modes.
\begin{itemize}
\item $\lambda_2=\lambda_3=\lambda_{23}=\Lambda_2$, $\lambda_{12}=\lambda_{13}=\Lambda_{12}$, $\lambda_1=\Lambda_1$
$$
H= -\Lambda_1 :\left(\vec{s}_1(0)\right)^2:  -\Lambda_2 :\left(\vec{s}_2(0)+\vec{s}_3(0)\right)^2:-
2\Lambda_{12}: \vec{s}_1(0)\cdot\left(\vec{s}_2(0)+\vec{s}_3(0)\right):
$$
\end{itemize}
This case corresponds to the two-stage Kondo effect. However, unlike conventional 2SK, the screening at the first stage is done by two orbital channels such a way that spin $S=1$ is screened first and $s=1/2$ is screened at the second stage.

The most general form of the low-energy FL Hamiltonian for the three-stage Kondo problem corresponding to the particle-hole symmetric limit of the three-orbital-level Anderson
model is given by $H=H_0 + H_\alpha+H_\phi+H_\Phi$ with $i,j=e,o1,o2$:
\begin{align}
H_0&{=}\phantom{-}\sum_{i\sigma}\int_\varepsilon \nu\left(\varepsilon + \varepsilon_\sigma^Z\right)
b^\dagger_{i\varepsilon\sigma}b^\pdag_{i\varepsilon\sigma}\nonumber\\
  H_\alpha& {=}{-}\sum_{i\sigma}
  \int_{\varepsilon_{1-2}} \frac{\alpha_{i}}{2 \pi}  
\left(\varepsilon_1+\varepsilon_2\right)
 \!  b^\dagger_{i\varepsilon_1\sigma}b^\pdag_{i\varepsilon_2\sigma}\! \nonumber
\\
H_\phi& {=}\phantom{-}\sum_{i} \int_{\varepsilon_{1-4}} \frac{\phi_{i}}{\pi\nu} 
: \! b^{\dagger}_{i\varepsilon_1\uparrow}b_{i\varepsilon_2\uparrow}
b^{\dagger}_{i\varepsilon_3\downarrow}b_{i\varepsilon_4\downarrow} \! :\nonumber
\\
H_\Phi&{{=}}{{-}}{\sum_{ij\sigma_{{1{-}4}}}}
{\int_{\varepsilon_{{1{{-}}4}}}}
{{\frac{\Phi_{ij}}{{2}{\pi}{\nu}}}}
{{:}b^{{\dagger}}_{{i\varepsilon_{1}\sigma_{1}}}{\boldsymbol{\tau}_{{{\sigma_{12}}}}}
{b_{{i\varepsilon_{2}\sigma_{2}}}} {b^{{\dagger}}_{j\varepsilon_{3}\sigma_{3}}{\boldsymbol{\tau}_{\sigma_{34}}}
b_{j\varepsilon_{4}\sigma_{4}}{:}}}{,}
\label{HFL4}
\end{align}
where $\alpha_i=\phi_i$ in accordance with Nozieres theory. The six-parametric strong coupling fixed point Hamiltonian (\ref{HFL4}) accounts for both elastic and inelastic processes. Finite temperature conductance
behaviour is controlled by three Kondo temperatures $T_{K}^i\propto \lambda_i^{-1}$ and three additional parameters ${\mathcal F}_{ij} \propto (\lambda_i-\lambda_{ij})(\lambda_j-\lambda_{ij})/(\lambda_i-\lambda_j)^2$ in full accordance with \cite{twoc}.

\end{widetext}
\end{document}